\documentclass[aps,prb,twocolumn,epsfig]{revtex4}  
\usepackage{epsfig}% Include figure files

\begin{document}

\title{Field-Driven Transitions in the Dipolar Pyrochlore Antiferromagnet Gd$_2$Ti$_2$O$_7$}
\author{Olivier C\'epas$^*$ and B. Sriram Shastry$^{*,\dagger}$}
\affiliation{
$*$ Department of Physics, Indian Institute of Science, Bangalore 560012, India. ,\\
$\dagger$ Department of Physics, University of California, Santa Cruz, 95064.}

\date{\today}

\begin{abstract} We present a mean-field theory for magnetic field
driven transitions in dipolar coupled gadolinium titanate
Gd$_2$Ti$_2$O$_7$ pyrochlore system.  Low temperature neutron
scattering yields a phase that can be regarded as a 8 sublattice
antiferromagnet, in which long-ranged ordered moments and fluctuating
moments coexist.  Our theory gives parameter regions where such a
phase is realized, and predicts several other phases, with transitions
amongst them driven by magnetic field as well as temperature. We find
several instances of {\em local} disorder parameters describing the
transitions.  \end{abstract}

\pacs{PACS numbers:}
\maketitle
\section{Introduction}

The pyrochlore S=7/2 Heisenberg spin system Gd$_2$Ti$_2$O$_7$ is
currently very popular.  In addition to magnetic ordering at 1
K\cite{Raju} which is not expected for a Heisenberg pyrochlore
system,\cite{Moessner} it displays magnetic field-driven phase
transitions that are intriguing, and have been ascribed to the
competition between dipole-dipole (d-d) interactions and the Zeeman
energy on frustrated {\em and} undistorted cubic
systems.\cite{Ramirez} The exchange energy has been argued to be
relatively small, due to the compact f-shell of Gd, but nevertheless
plays an important role in lifting the degeneracy between various
possibilities, as we show in this work.

Earlier examples of similarly low energy scales are known, but on the
diamond lattice\cite{White} and the gadolinium gallium
garnet,\cite{GGG} thus lacking the complexity of the
pyrochlores. Theoretical interest in d-d coupled systems is quite
old,\cite{luttinger} focussing on its long ranged nature.

Gd$_2$Ti$_2$O$_7$ is intriguing, first by the magnetic order it
displays at low temperatures and zero magnetic field, as observed by
neutron scattering.\cite{Champion} The magnetic order has been
assigned a propagation vector $\bf{q}=\{\pi,\pi,\pi \}\equiv
\mbox{\boldmath$\pi$}$ at 50 mK, and the interstitial
moments\cite{interstitial} are zero on average, thus showing a
partially-ordered state despite the very low
temperature.\cite{Champion} This structure is consistent with the
correlations observed at higher temperatures in ESR.\cite{Hassan} This
turns out to be in direct conflict with previous theoretical works
that assumed a nearest neighbor Heisenberg system with d-d
interactions.  The expansion of the free-energy to quadratic order in
the order parameter near the ordering temperature, $T_N$, tells us
that all the states with a propagation vector along the $(111)$
direction become unstable simultaneously.\cite{Raju} Such a degeneracy
is in fact not quite exact; refined numerics and analytical work show
that the $\textbf{\mbox{q}}= \mbox{\boldmath$\pi$}$ state is very
slightly preferred.\cite{Huse} Nevertheless, below $T_N$ the fourth
order term in the free energy expansion becomes important and favors a
$q=0$ state,\cite{Palmer} which has been confirmed by a real-space
mean-field theory employing a four sublattice decomposition of the
pyrochlore lattice.\cite{Ramirez} From these works, however, it is not
clear as to how this $\textbf{\mbox{q}}= \mbox{\boldmath$\pi$}$ state
can be realistically and {\em robustly} realized.  We find that this
requires going beyond the nearest neighbor exchange model studied
previously.

The second particularly interesting feature of gadolinium titanate is
the presence of magnetic-field-driven phase transitions.\cite{Ramirez}
For cubic systems one expects exotic phases connected by continuous
transitions, as compared to well studied first order spin-flop type
transition on uniaxially distorted systems, such as MnF$_2$. Although
the four sublattice mean-field theory is in conflict with the neutron
scattering results, it gives such field-driven transitions with critical fields
in rough agreement with those observed on a powder
sample.\cite{Ramirez}

The aim of the present article is to show that a model that includes
exchange energies beyond the nearest neighbors not only explains the
magnetic structure observed by neutron scattering at low temperatures
but also exhibits magnetic-field-driven transitions whose features can
be checked by further experimentation. We present the results of the
expansion of the free-energy near $T_N$ together with a low
temperature mean-field theory using ``hard-spins" (that describes the
fixed length constraint well) that is, hence, valid at all
temperatures, although limited by the sublattice structure imposed at
the outset. The earlier analysis\cite{Ramirez} imposed a 4 sublattice
order, here we go to the next level of description, namely a 8
sublattice order, in order to accomodate the $\textbf{\mbox{q}}=
\mbox{\boldmath$\pi$}$ state. This allows us to find other stable
phases.

\begin{figure}[htbp]
\vspace{-0.2cm}
\centerline{
 \psfig{file=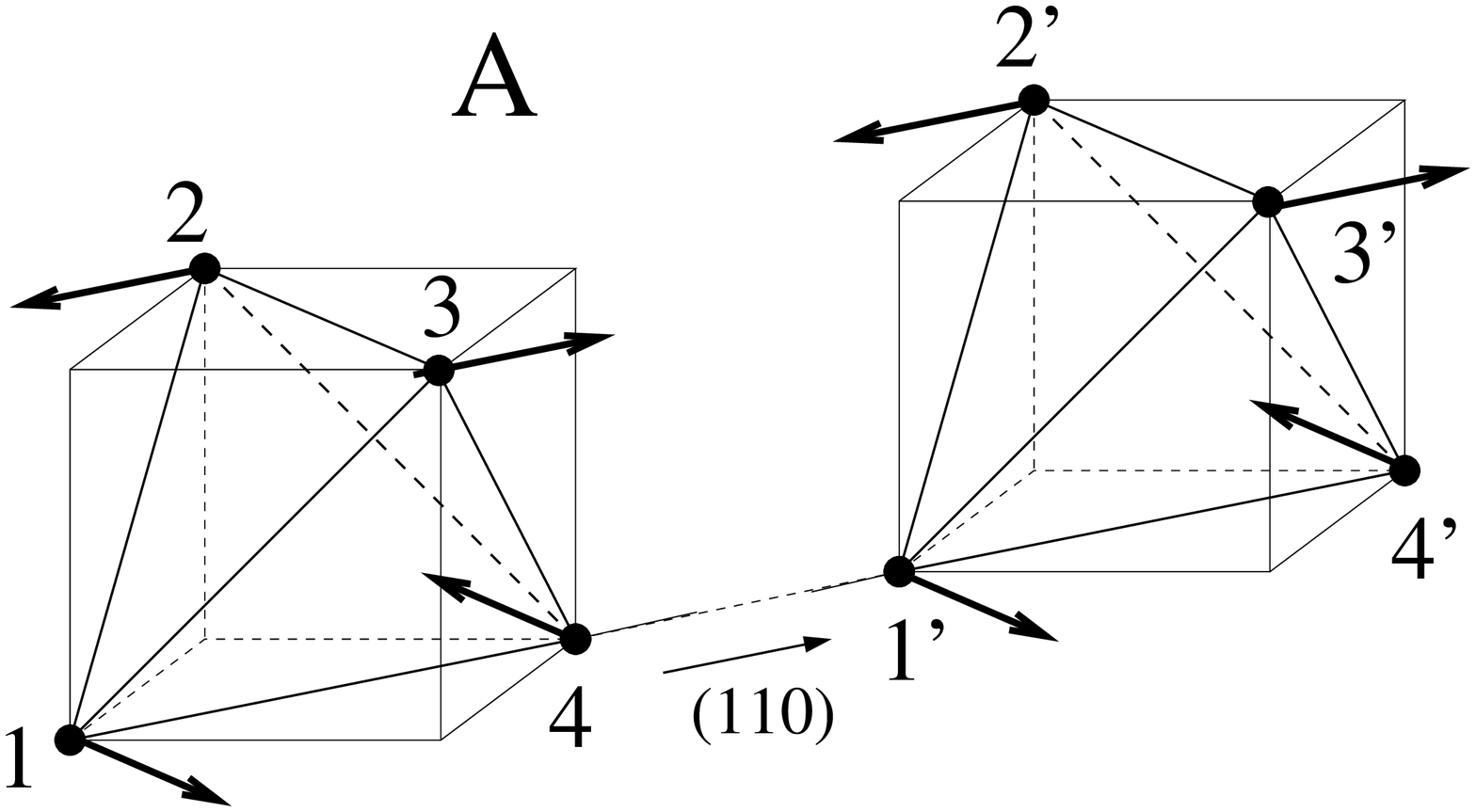,width=4.5cm,angle=0}
\hspace{0.6cm} \psfig{file=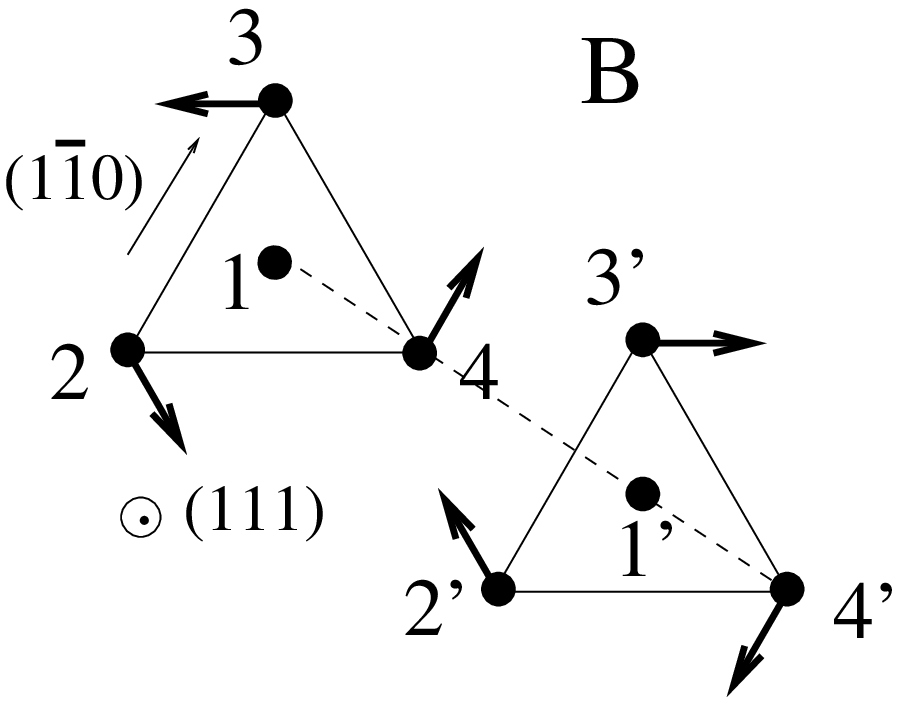,width=3.5cm,angle=0}}
\centerline{
 \psfig{file=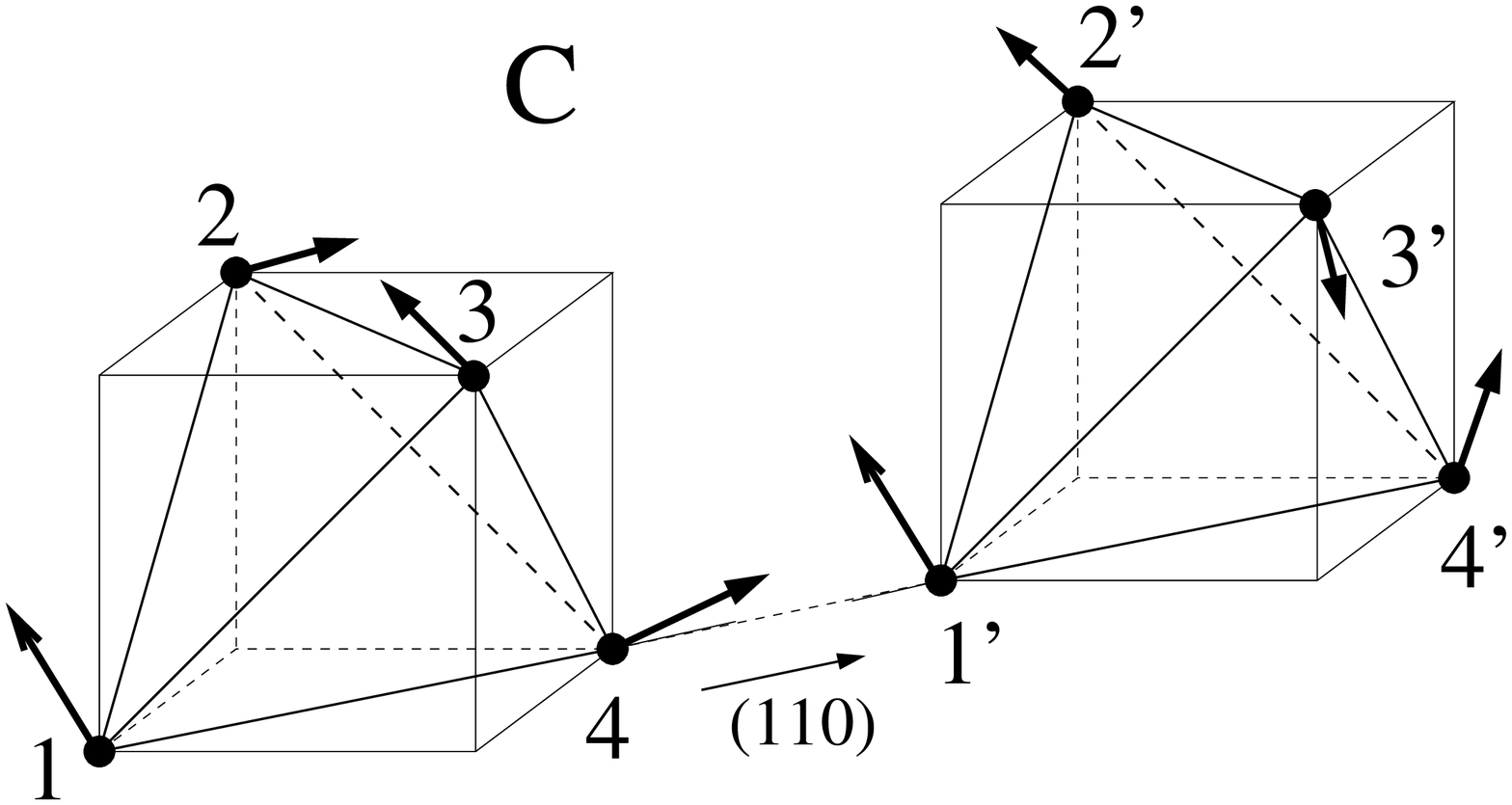,width=4.5cm,angle=0} \hspace{0.6cm}
\psfig{file=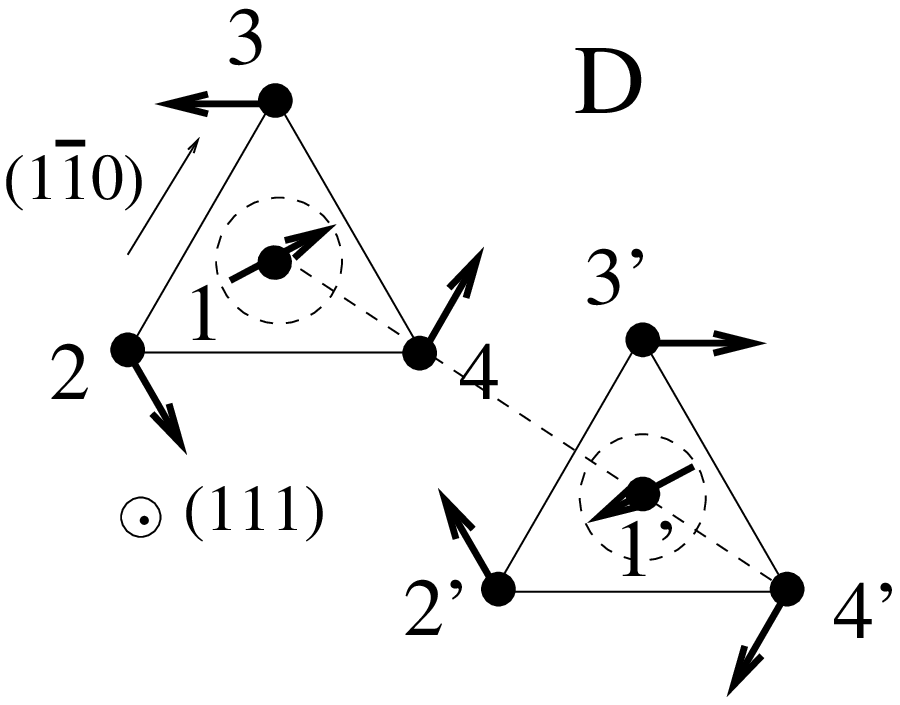,width=3.5cm,angle=0}} \vspace{0.2cm}
\caption{On the left, the eight sublattices used in the mean-field
calculation.  The A state (q=0) is shown ( six fold degenerate). On
the right, the spins of the B state (degeneracy 2) belong to the
Kagom\'e planes; the sublattices 1 and 1' have no magnetic moment in
average. The D state is identical to B except that the magnetizations
of the sublattices 1 and 1' are finite and opposite from one Kagom\'e
plane to the next (continuous degeneracy). C does not have any simple
description (see text).}  \label{states} \end{figure}

We begin by summarizing the results of the ``hard-spin" mean-field
theory.  The 8-sublattice system consists of two tetrahedra (see
Fig. \ref{states}), the second one being obtained by a translation of
the first one along the primitive vector $(110)$.  There are 4 other
such equivalent directions that would give equivalent results, in
particular with regards to the propagation vector
$\mbox{\boldmath$\pi$}$. Without loss of generality, we consider only
one of them in the following, namely $(111)$.

\section{Zero magnetic field phases}

The pyrochlore lattice sustains various phases. We have alluded to the
$\textbf{\mbox{q}}=\textbf{\mbox{0}}$ phase previously, that we denote A
in the present work. It has a 6-fold degeneracy according to the three
planes of the cubic structure and the time-reversal symmetry.\cite{Palmer,Ramirez} In addition there are three other phases that
we call B, C and D, that break translation invariance.

At finite temperatures, and for different (exchange) parameters, we
find a distinct phase B, that can be described as a
$\textbf{\mbox{q}}= \mbox{\boldmath$\pi$}$ state.  In this phase, the
interstitial sites have, on average, no magnetic moment. The spins of
the Kagom\'e planes, on the other hand, are ordered in a $120$ degree
structure; each of them being parallel to the opposite edge in order
to minimize the d-d interactions. This is the phase that has been
found experimentally in Gd$_2$Ti$_2$O$_7$.

There is also a phase that we call C which does not have a simple
description: the interstitial spins ($1-1'$) are parallel to each
other and the other spins break translational symmetry and are not
coplanar. There are 12 degenerate such states. It seems plausible that
the C-type phases are in reality the `projection' onto the
8-sublattice of incommensurate states.

In addition to these phases which all have a finite degeneracy, we
have found a phase, D, which is particularly interesting in that it
has a continuous degeneracy. We emphasize here that we actually start
from a strongly frustrated system, whose huge degeneracy is lifted by
the d-d interactions.  It is therefore surprising to obtain a
continuously degenerate ground state in a wide range of parameters
when further neighbor interactions are switched on. In the D phase,
the spins of the Kagom\'e planes are identical to those of the B
phase. The interstitial moments, however, are finite. They may point in
any direction in a plane parallel to the Kagom\'e planes, giving a
large degeneracy, and they are antiparallel to each other from plane
to plane.  This is actually the zero temperature analogue of the B
phase, but occurs at low temperatures.

\begin{figure}[htbp]
\vspace{-0.5cm}
\centerline{
 \psfig{file=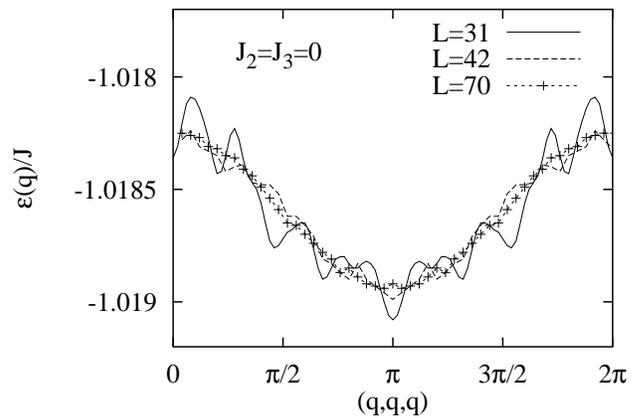,width=6cm,angle=-90}} 
\vspace{0.2cm}
\caption{Quasi-degeneracy of the lowest eigenvalue as function of  $\textbf{\mbox{q}}$ along $(111)$ for $J_2=J_3=0$ (see the horizontal scale). 
When $L$ (the number of neighbors in the $(100)$ direction included in the dipolar sums) is increased, the curves converge to a smooth curve which minimum is at, or very close to $\textbf{\mbox{q}}= \mbox{\boldmath$\pi$}$.}
\label{convergence}
\end{figure}

\begin{figure}[htbp]
\vspace{-0.5cm}
\centerline{
 \psfig{file=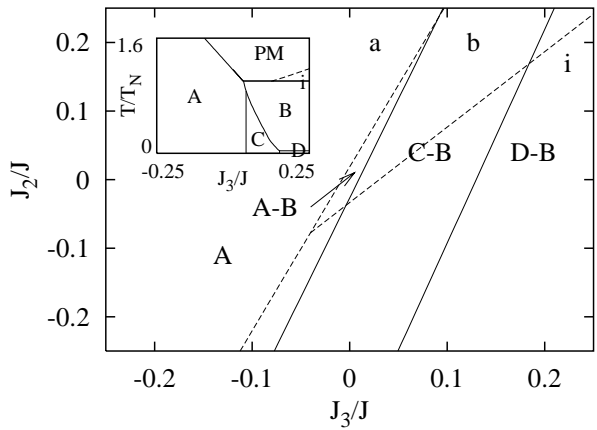,width=6cm,angle=-90}} 
\vspace{0.1cm}
\caption{Phase
diagrams showing the first unstable modes of the paramagnetic phase at
$T_N$ (dashed line, small letters, `` i '' for incommensurate) and
from the $8$-sublattice mean-field theory (solid line, capital
letters).  In the inset, the phase diagram as a function of
temperature is shown for $J_2=0.1J$.  $T_N$ is the temperature to
enter the B phase at intermediate $J_3$. $J=0.4$ K.}  \label{phasediagram}
\end{figure}

To find those phases, we have considered the Hamiltonian defined on
the pyrochlore lattice 
\begin{eqnarray} {\cal H} &=& \frac{1}{2}
\sum_{ij} J_{ij} \textbf{\mbox{S}}_i \cdot
 \textbf{\mbox{S}}_j - g \mu_B \sum_i \textbf{\mbox{S}}_i \cdot
 \textbf{\mbox{H}} \nonumber \\
& &  + (g \mu_B)^2 \sum_{ij} \left(
 \frac{\textbf{\mbox{S}}_i \cdot \textbf{\mbox{S}}_j}{r_{ij}^3} - 3
 \frac{(\textbf{\mbox{r}}_{ij} \cdot \textbf{\mbox{S}}_i)
 (\textbf{\mbox{r}}_{ij} \cdot \textbf{\mbox{S}}_j)}{r_{ij}^5} \right)
\end{eqnarray} where $ \textbf{\mbox{S}}_i$ is a quantum spin operator
for spins $S=7/2$ on site $i$. $J_{ij}$ is the Heisenberg exchange
between the neighbors: we are in fact considering the first neighbors
at a distance $a_0=3.59 \AA$ ($J$), the second ($J_2$, distance
$\sqrt{3}a_0$) and third neighbors ($J_3$, distance $2a_0$).  We note
that there are in fact two types of third neighbor exchanges according
to the crystal structure. The difference of them being another small
parameter, we will neglect it in the following.  Let us first describe
the ``k-space" mean-field theory.  We have studied the stability of
the paramagnetic state towards a state with a modulation vector $k$,
extending Ref.\onlinecite{Raju} to include $J_2$ and $J_3$.  By
expanding the free-energy near $T_N$, we get a $12 \times 12$ matrix,
its lowest eigenvalue determines $T_N$ and the corresponding
eigenvector describes the $k$-modulation of the state. We find that
the degeneracy along $(111)$\cite{Raju,Palmer} for $J_2=J_3=0$ is
actually weakly lifted when the dipolar sums include a larger number
of neighbors compared with previous works, as previously
noticed.\cite{Huse} The selected mode is at, or very close to,
$\textbf{\mbox{q}}= \mbox{\boldmath$\pi$}$ (see
Fig. \ref{convergence}). New numerical work using the Ewald summation
technique has indeed clearly shown that this mode is
selected.\cite{Gingras}
Nevertheless in order to robustly lift what remains a
quasi-degeneracy, it is important to include $J_2$ and $J_3$.  As soon
as $J_2<0$, $\textbf{\mbox{q}}= \mbox{\boldmath$\pi$}$ is robustly
selected. The corresponding eigenvector tells us that it corresponds
to the B state.  There are two other regions of the phase diagram
where the A phase and an incommensurate phase are preferred.  The
phase diagram of the first instability is given in dashed lines in
Fig. \ref{phasediagram}.
Since the above approach can only give the first instability that is
encountered on cooling, i.e. at $T_N$, nothing is known about the
lower temperature behavior of the system. We therefore proceed to a
real-space mean-field theory following Ref. \onlinecite{Ramirez}, but
enlarging the unit cell to 8 sublattices, as already specified.

Within the mean-field theory, the magnetizations of the sublattices
are given by $\langle \textbf{\mbox{S}}_a \rangle= S
\hat{\textbf{\mbox{h}}}_a {\cal B}_S(\beta h_a S)$ with the
definitions of the local fields, $h_a^{\alpha} \equiv g \mu_B
H^{\alpha} - \sum_{b \beta} J_{ab}^{\alpha \beta} \langle S_b^{\beta}
\rangle $, where ${\cal B}_S$ is the Brillouin function, and $\beta$
is the inverse temperature.  $J_{ab}^{\alpha \beta}$ couples the
sublattices $a$ and $b$, running from 1 to 8.  A straightforward way
of solving the problem consists of iterating numerically these
self-consistent equations, starting from many random configurations
and selecting the set of states with the lowest free-energy. This
leads to the phase diagram given in Fig. \ref{phasediagram} (solid
lines). The phases A, C and D described above are found at zero
temperature in quite a narrow and physically reachable region of
parameters.  However, there is no sign at zero temperature of the B
phase found in the mode analysis.

In Fig. \ref{phasediagram}, we give the labels for the sequence of
phases on increasing T prior to reaching the paramagnetic
phase. Details of the transition temperatures for the $8$-sublattice
problem are given in the inset of Fig. \ref{phasediagram} for a given
$J_2=0.1J$ (solid lines).  For $J_2=J_3=0$, as another example, the A
phase is definitely chosen below $T_N$, in agreement with
Refs.\onlinecite{Palmer,Ramirez}.  The transition to the B state at
$T_N$, as suggested in the mode analysis given above
(fig. \ref{convergence}), is in fact immediately followed by a
transition to the A phase when the temperature is decreased (so that
on the scale of fig. \ref{pd0}, the B phase is indistinguishable at
the PM-A transition). To robustly select the B phase in a wide range
of temperatures (as experimentally observed) it is necessary to
include other interactions, the simplest assumption being the next
nearest Heisenberg couplings. Indeed when $J_3$ is increased or $J_2$
decreased, the region where we have the A-B succession of phases
widens (see inset of Fig. \ref{phasediagram}), before the onset of the
other phases. For these other phases, the C-B transition line appears
to be first-order, whereas D-B is second-order.

We can now compare in more details these results with those of the
$k$-space mean-field theory, i.e. the stability analysis of the
paramagnetic state.  For large portions of the phase diagram (where
the last capital letter coincides with the small letter), the two
approaches give the same result. There is, however, a region of
parameters (e.g. large $J_3$) where the first unstable mode is
incommensurate, while for the same parameters we find a transition
from the paramagnetic state to the B state in the 8-sublattice
calculation. The 8-sublattice calculation can not capture the first
transition to the incommensurate state because of the sublattice
decomposition. On the other hand, such a calculation respects the
fixed length constraint of the spins at low temperatures. Within this
point of view, two scenarios may take place when the temperature is
lowered. In both of them there is first a transition from the
paramagnetic state to the incommensurate state.  Then the
incommensurate state may exist down to zero temperature and the
occurrence of the B state is a pure artifact of the sublattice
calculation. Alternatively, the B state \textit{is} stabilized at low
temperature. In this case there must be a phase transition between the
incommensurate state and the B state when the temperature is lowered.
In order to rule out one scenario, one would need to study the
stability of the phases at low temperature bypassing the limitations
of the sublattice decomposition.

\section{Magnetic-Field-Driven Transitions}

\begin{figure}[htbp] \vspace{-0.5cm} \hspace{-0.5cm}
\psfig{file=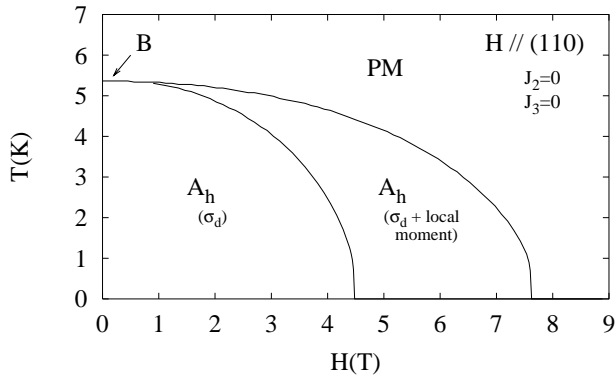,width=5.2cm,angle=-90} \vspace{0.1cm}
\caption{Phase diagram (H,T) for a magnetic field parallel to $(110)$,
showing the broken symmetries (in brackets). A$_{\mbox{h}}$ represents
the A phase under field. The B phase occurs in a very narrow
region near $T_N$ at $H=0$ (see text).} \label{pd0}
\end{figure}

A magnetic field is an interesting probe of these phases.  We
highlight here the main results on all four phases since we believe
that the A, C and D phases might be useful for other materials. The A
phase gives rise to multiple phase transitions, as previously reported
for T=0.\cite{Ramirez} We note, however, several new elements arising
from the effect of finite temperatures.  In Fig. \ref{pd0}, we give
the example of the complete phase diagram when the field is along
$(110)$. For small enough fields, a raise in temperature drives the
system through two phase transitions, a result which also holds for
the fields along the other crystallographic directions, $(100)$ and
$(111)$, with, in the latter case, a first transition which is weakly
first-order.

\begin{figure}[htbp] \vspace{-0.5cm} \hspace{-0.5cm}
\psfig{file=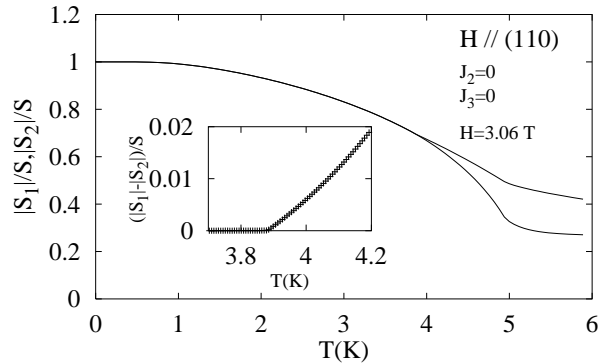,width=5.2cm,angle=-90} \vspace{0.1cm}
\caption{Local magnetic moments averaged by the thermal fluctuations
as function of temperature. Above the first transition, we see a difference
in the moments of the two sublattices that are inequivalent because of
the field.  The difference is a \textit{disorder} parameter for one
transition, with a critical exponent numerically close to $4/3$.}
\label{pddisorder}
\end{figure}

Although the two successive phases A$_{\mbox{h}}$ for a field along
$(110)$ are separated by a transition line (see Fig. \ref{pd0}), they
can not be distinguished by a different \textit{geometrical} broken
symmetry. They basically bear the same $\sigma_d$ broken symmetry, but
none of the other geometrical symmetries is broken in either phase.
The effect of thermal fluctuations turns out to be crucial in
distinguishing them.  Indeed, at any finite temperatures, this
transition is associated with the occurence of a \textit{disorder}
parameter - a quantity which is zero in the ordered phase and non-zero
in the disordered phase, which we now describe.  When we look at the
value of the magnetic moment (smaller than $S$ as soon as $T>0$), it
appears that in the ordered phase all the magnetic moments of the four
sublattices are the same, while they are different above the first
transition (Fig. \ref{pddisorder}).  We can thus define a {\em
disorder parameter} by the difference of two of these moments (inset
of Fig. \ref{pddisorder}).  Moreover, we have found that the disorder
parameter has a critical exponent $\sim 4/3$. This is in 
contrast with the other linear order parameters which follow the
usual $1/2$ exponent of the Landau theory. One remarkable feature to
note is that our disorder parameters are {\em local}. This is quite
unlike familiar disorder parameters that arise in theories with
duality that lead naturaly to highly nonlocal disorder parameters,
such as are known for the 2-d Ising model.

We note that these transitions are quite different from the classical
spin-flop transitions of the uniaxial antiferromagnets. In addition to
those unconventional mean-field exponents, the response of the spins
themselves is worth mentioning.  When a field along $(100)$ is
increased, for instance, the spins 1 and 2 of the A phase (or 3 and 4)
which are perpendicular to each other at zero field, remain exactly
perpendicular while the sum of both spins twist towards the field.
This is not obvious and in particular can not be understood by considering partial couples, since it is a real 4-sublattice-coupled system.

\begin{figure}[htbp] \vspace{-0.2cm}\centerline{
\psfig{file=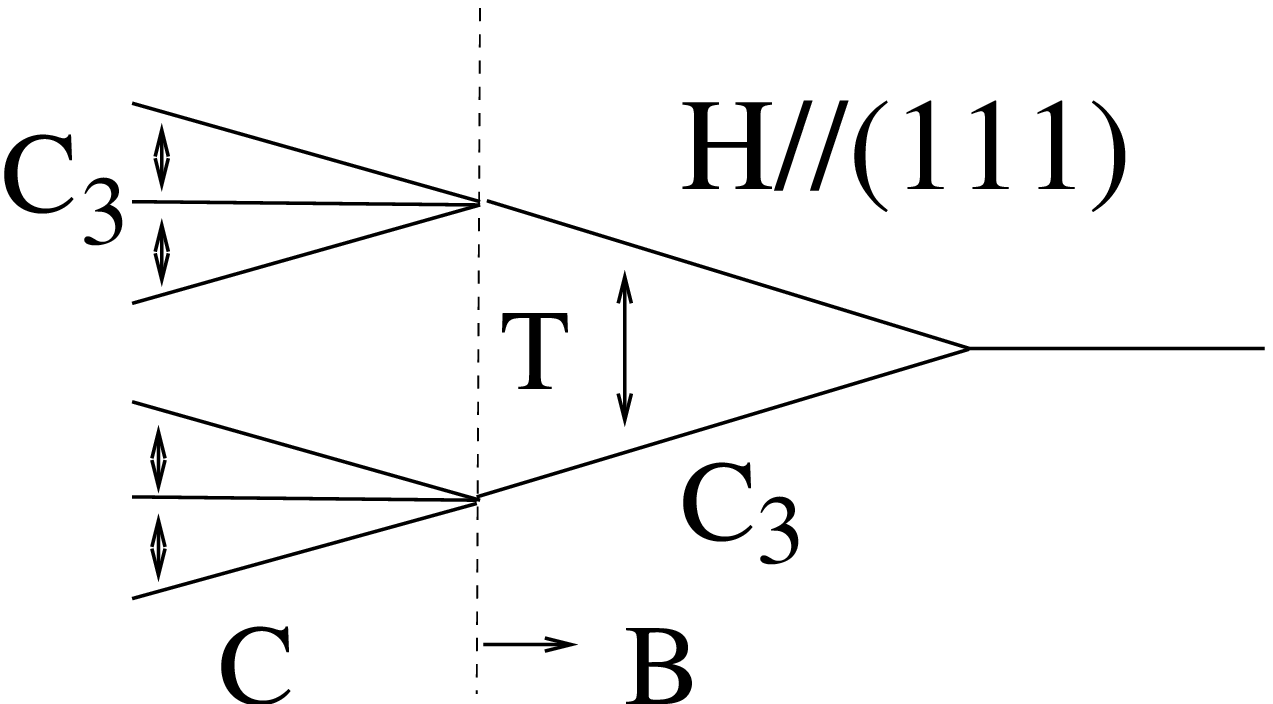,width=3.5cm,angle=0} \hspace{0.8cm}
\psfig{file=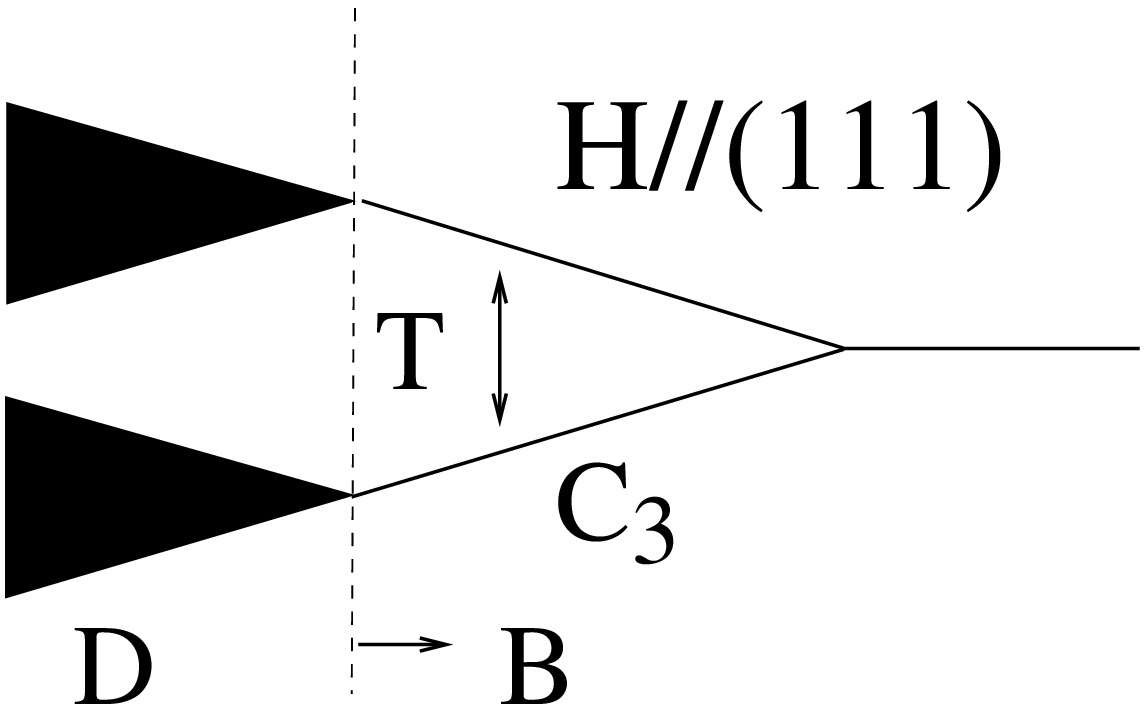,width=3.0cm,angle=0}} \centerline{
\psfig{file=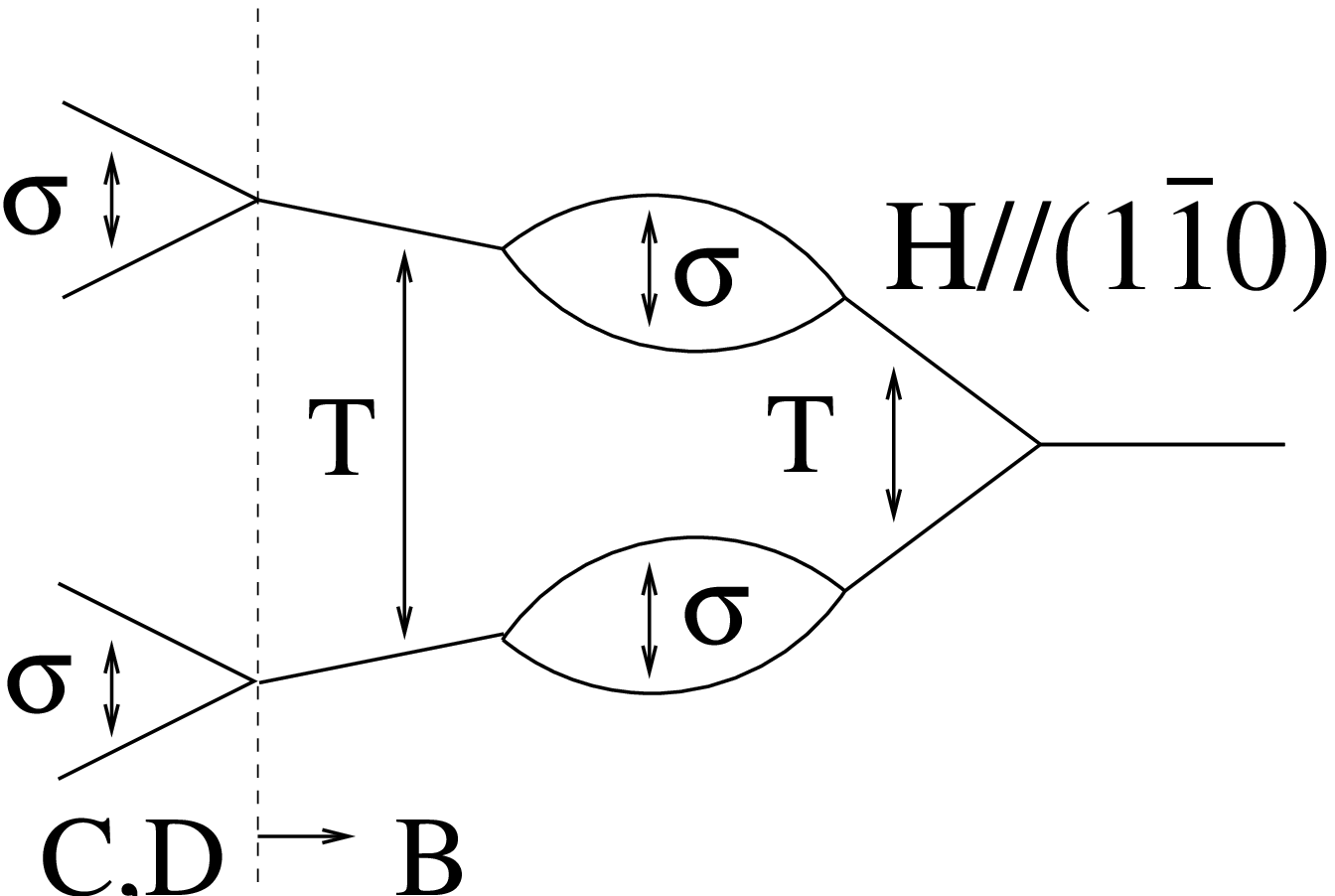,width=3.5cm,angle=0}} \centerline{
\psfig{file=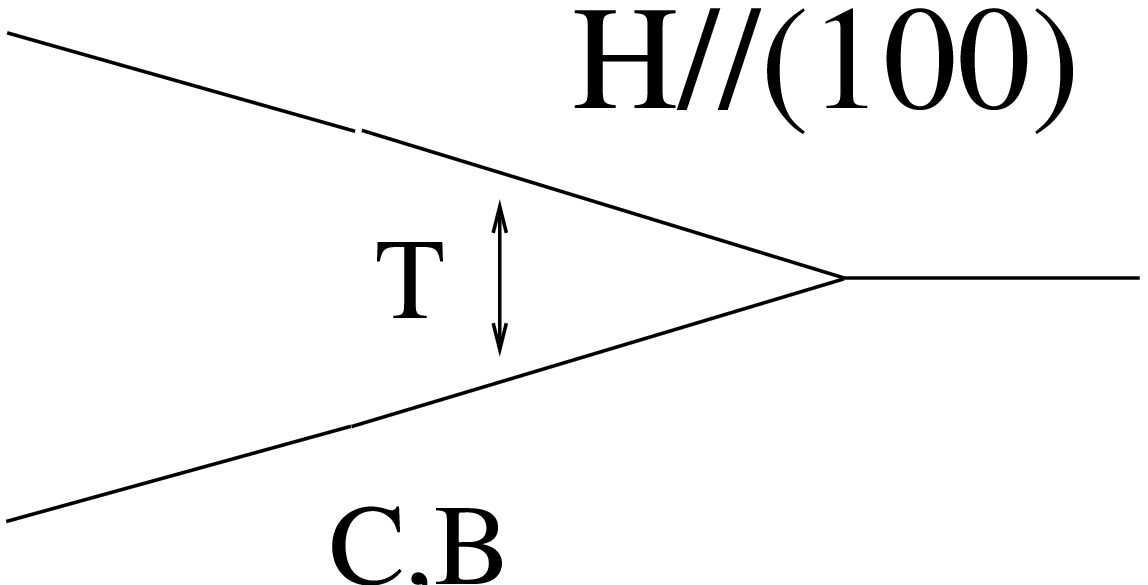,width=3.0cm,angle=0} \hspace{0.8cm}
\psfig{file=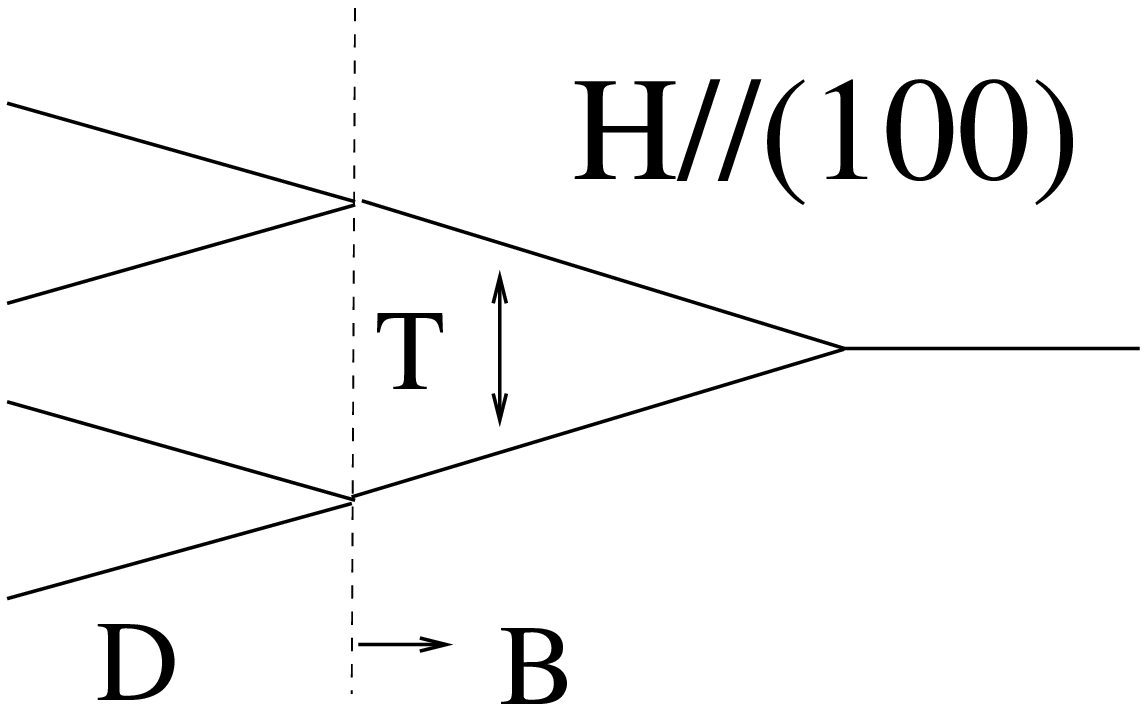,width=3.0cm,angle=0}} \vspace{0.5cm}
\caption{Evolution of the degeneracies as a function of an external
magnetic field along the main crystallographic directions. The broken
symmetries and the symmetries that connect the distinct states are
also given ($T$ is the translation operation, $\sigma$ a mirror plane,
and $C_3$ a three-fold rotation). Dark shading indicates continuous
degeneracy of the state.}  \label{diagrams} \end{figure}

The effect of a magnetic field on the B-phase is to induce back a net
magnetic moment on the previously zero moment interstitial sites. At
higher fields along the $(111)$ or $(100)$ directions, there is a
unique transition to the paramagnetic phase with a merging of the two
degenerate states (Fig. \ref{diagrams}).  For the $(1\underline{1}0)$
direction, however, there is a reentrance of a less symmetric phase
when the magnetic field is increased. This phase is 4-fold degenerate
and breaks the mirror plane symmetry. At larger fields the 2-fold
degenerate state is recovered before reaching the transition to the
paramagnetic state. In total, there are three main distinct
transitions when the field is increased from zero because the critical
field of the paramagnetic transition depends very weakly upon the
direction of the field.

Regarding the C or D phases, we have to describe what happens to the
interstitial moments before the system enters the B-phase and
undergoes the transitions described above. For the D-phase, the
continuous degeneracy is preserved with a field along $(111)$, until
the B-phase is reached.  For the other directions, the degeneracy is
lifted at an infinitesimal field by a spin-flop transition for the
interstitial moments. For the C-phase, the situation is very similar
(see Fig. \ref{diagrams}).

\section{Comparison with experiments}

We now compare our theoretical calculations with experimental results
on Gd$_2$Ti$_2$O$_7$. The success of this approach is to predict the
existence of the $\textbf{\mbox{q}}= \mbox{\boldmath$\pi$}$ phase in a
wide range of temperatures and parameters. Although it is not the
ground state of the problem (since only A, C and D phases are obtained
at zero temperature), the B phase may be stabilised at very low
temperatures, as observed experimentally. This is obtained with 
parameters $J_2$ and $J_3$ typically of the order of $0.1J$ which are
physically reasonable.  Next nearest neighbor Heisenberg interactions
may indeed take place in these materials, especially
if we think of an exchange mechanism in terms of a magnetic exchange
between the f-electrons and the more extended d electrons which may
carry the spin polarization at distances larger than the nearest
neighbors.\cite{Bouzerar}
\begin{figure}[htbp] \vspace{-0.5cm} \centerline{
\psfig{file=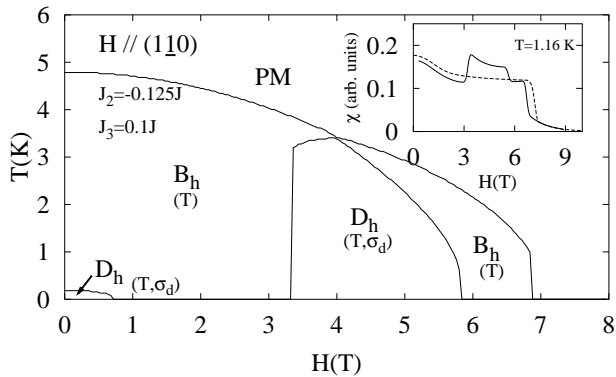,width=5.2cm,angle=-90}} \vspace{0.2cm}
\caption{Phase diagram (H,T) for a magnetic field parallel to
$(1\underline{1}0)$, showing the various phases and the broken
symmetries (in brackets). D$_{\mbox{h}}$ represents the D phase under
field. Insert: Spin susceptibility for $H
\parallel (1\underline{1}0)$, convoluted with a gaussian function
(solid line) and powder-averaged susceptibility (dashed line).}  
\label{pd} \end{figure}
We give the phase diagram for typical parameters in Fig. \ref{pd},
which reproduces the number of field-driven transitions observed
experimentally.  We have two free parameters, $J_2/J$ and $J_3/J$,
chosen typically as $-0.125,0.1$, and $J$ is then fit to the measured
high-temperature susceptibility. The values of the critical fields
measured at low temperatures ($3$T, $6$T and $7$T) are then well
reproduced. Spin susceptibilities are given in the inset of
Fig. \ref{pd}. Note that the powder-averaging susceptibility is almost
featureless and does not account for the complexity of the
transitions.

Also interesting are the ESR results in the paramagnetic phase
.\cite{Hassan} In addition to anisotropic properties, the ESR signal
shows two lines, revealing the existence of two inequivalent sites,
though since no structural distortions have been reported, the four
sites of the unit-cell should be equivalent.  We note for example that
distortions that would strengthen the magnetic interactions in the
Kagom\'e planes would naturally help to select the B phase.  Even in
the absence of distortions, the applied magnetic field (which is
necessary to perform the ESR studies) breaks some symmetries of the
lattice. In this respect, and although the sublattices are equivalent
at zero magnetic field, we have found different averaged magnetic
moments on two different sites under magnetic field. Therefore, we
naturally expect two lines to appear in the high-temperature ESR
spectrum. Note in particular that the difference of magnetic moments
is precisely what we found to be a disorder parameter in one
transition at low temperature (fig. \ref{pddisorder}). In any case, it
is an interesting example where ESR would not probe the properties of
the zero-field state. This is of course expected from symmetry
considerations, but we have found here a simple example of a system
where it does occur.

The phase diagram as a function of the model parameters
(fig. \ref{phasediagram}) may be relevant to other compounds as
well. The chemical replacement (of Ti by Sn for instance), or an
external pressure could well drive the system into either of the other
phases because of the modifications in the magnetic exchange paths,
and hence the relative strengths of $J$, $J_2$, and $J_3$.  A
different behavior has indeed been found experimentally in
Gd$_2$Sn$_2$O$_7$ at low temperatures.\cite{Bertin} Although this
system orders at a similar temperature, the moments were suggested to
be perpendicular to the local $(111)$ directions on the basis of
M\"ossbauer measurements.\cite{Bertin} Such a magnetic ordering, if
confirmed by neutron scattering, would be more compatible with
the A phase.

\section{Conclusion}

In summary, we have found several phases and phase transitions in the
dipolar pyrochlore lattice by taking into account the magnetic
exchange beyond the nearest neighbour Heisenberg coupling. One of them
is precisely the B phase of Gd$_2$Ti$_2$O$_7$ seen in neutron
scattering.\cite{Champion} We find that it is not the ground-state,
but a finite temperature state. We obtain, as ground states, the A, C
or the degenerate D phase.  When a magnetic field is applied, we find
that the B-phase undergoes three transitions, as appeared in specific
heat measurements. We have predicted the corresponding magnetic
structures and the broken symmetry phases that could be checked by
neutron diffraction experiments. Sound velocity and absorption
studies, as well as calorimetric studies with aligned crystals in
magnetic fields should shed light on the nature of the phases
predicted here.

\acknowledgements

We thank Art Ramirez and David Huse for stimulating discussions and
for constructive suggestions, and R. Karan for numerical help. We
acknowledge support from an Indo-French grant IFCPAR/2404.1.


\begin{references}
\bibitem{interstitial} The pyrochlore lattice can be regarded as a set of Kagom\'e planes connected by ``interstitial'' gadolinium sites that link the planes.
\end{references}
\end{document}